\documentclass[letterpaper, 11pt]{article}

\usepackage{mathtools}
\usepackage{array}
\usepackage{enumitem}
\usepackage{amsmath,amsfonts,amssymb,bm}
\usepackage{titlesec}
\usepackage{pgfgantt,rotating}
\usepackage{float}
\usepackage{graphicx}
\usepackage{sidecap}
\usepackage{wrapfig}
\usepackage{bold-extra}
\usepackage{authblk}
\usepackage{appendix}
\usepackage{makecell}
\usepackage{dirtytalk}
\usepackage{placeins}
\usepackage{subfigure}
\usepackage[export]{adjustbox}
\usepackage{cite}
\usepackage{graphicx}
\usepackage{subfigure}
\usepackage{epstopdf}
\usepackage[margin=1in]{geometry}
\usepackage{algorithm}
\usepackage{algpseudocode}
\usepackage{amsmath}
\usepackage{booktabs}
\usepackage{multirow} 
\newcommand{\beq}{\begin{equation}}
\newcommand{\eeq}{\end{equation}}
\newcommand{\bgqar}{\begin{eqnarray}}
\newcommand{\enqar}{\end{eqnarray}}
\newcommand{\bgqarn}{\begin{eqnarray*}}
\newcommand{\enqarn}{\end{eqnarray*}}
\newcommand{\bgary}{\begin{array}}
\newcommand{\enary}{\end{array}}

\graphicspath{{Figures/}}

\title{Operator-Theoretic and physics-guided Sequence Modeling of Lithium-Ion Battery Voltage Dynamics}
\author{Khalid Mahmud Labib$^{1}$, Inayat Rasool$^{1}$, Shabbir Ahmed$^{1{\dagger}}$}

\affil{\small $^{1}$Dynamical System and Signals Lab (DSSL)\\ Department of Mechanical Engineering\\
South Dakota State University, Brookings, SD }

\begin{document}

\maketitle

%------------------------------------------------------------------------------------------------------------
% Abstract
%------------------------------------------------------------------------------------------------------------
\begin{abstract}

Lithium-ion batteries exhibit nonlinear voltage dynamics across varying operating conditions and aging states, making accurate modeling essential for estimation, control, and health monitoring. This work compares two data-driven frameworks for modeling voltage responses from hybrid pulse power characterization (HPPC) measurements: an operator-theoretic model based on Dynamic Mode Decomposition with control (DMDc), and a physics-guided transformer-based sequence model. In the DMDc framework, delay-embedded snapshots of terminal voltage and current are used to identify system matrices directly from measurement data, yielding an interpretable state-space model for recursive prediction. In parallel, a modified PatchTST architecture is developed in which terminal voltage is decomposed into an analytically computed open-circuit-voltage (OCV) component and a learned dynamic residual, with a future-current fusion pathway tailored to the prescribed HPPC current profile. Experimental results on a 30 Ah lithium-ion cell show that, although both models capture the sharp transient pulse dynamics, DMDc achieves lower prediction error and greater robustness to cell degradation under the present limited data regime, while the transformer captures qualitatively similar dynamics with greater architectural flexibility. These results highlight the advantages of operator-theoretic models in interpretability, computational efficiency, and robustness, while indicating the promise of physics-guided transformer models when larger and more diverse datasets are available.
\end{abstract}

\section{Introduction}
%%%%%%%%%%%%%%%%%%%%%%%%%%%%%%%%%%%%%%%%%%%%%%%%%%%%%%%%%%%%%%%%%%%%%%%%%%%%%%%%
Lithium-ion batteries play a vital role in sectors such as electric vehicles, aircraft power systems, consumer electronics, and large-scale energy storage technologies \cite{FARHANGDOUST2023} due to their high energy density, high power capability, and long service life \cite{ahmed2025autoregressive}. The safe and efficient operation of these systems relies heavily on accurate prediction and monitoring capability of battery management system (BMS)\cite{Xiong2018,Wang2020}.Consequently, BMS requires reliable models to predict voltage dynamics, estimate internal states, and track degradation over time.

Battery modeling approaches are commonly categorized into physics-based models, equivalent circuit models (ECMs), and data-driven methods. Physics-based models, such as the pseudo-two-dimensional (P2D) formulation, can represent electrochemical processes and degradation behavior in considerable detail \cite{Lee2012,sancarlos2021rom,li2013reduced}, but they require the numerical solution of coupled partial differential equations and therefore incur high computational cost. ECMs provide a simpler alternative and are widely used in real-time applications because of their low computational burden \cite{lai2018comparative,quelin2023coupling,tran2021comprehensive,petri2023state}. However, their predictive performance depends strongly on the assumed circuit topology and parameter identification, and their ability to represent degradation-dependent dynamics is often limited.

These limitations have motivated increasing interest in data-driven approaches that infer battery behavior directly from measurements. Methods such as sparse identification of nonlinear dynamics (SINDy) and nonlinear autoregressive models with exogenous inputs (NARX) have been used to recover governing relations from data \cite{kaiser2018sparse,rudy2017data,piroddi2003identification,ahmed2021stochastic}. However, such methods are often sensitive to noise and do not naturally yield explicit state-space models for control, filtering, or modal analysis. In this context, Dynamic Mode Decomposition with control (DMDc) is attractive because it combines singular value decomposition with a linear controlled evolution model to identify operators that advance measurement snapshots forward in time while accounting for known inputs \cite{proctor2016dynamic,muld2012flow,pan2011dynamical}. When combined with delay-embedded observables, DMDc can provide an interpretable linear representation of nonlinear battery dynamics directly from voltage and current measurements.

At the same time, deep learning models have become increasingly attractive for battery forecasting because of their ability to approximate complex nonlinear input-output relationships. Sequence models based on attention mechanisms are particularly relevant for long-horizon forecasting, and recent architectures such as PatchTST improve scalability by representing temporal segments as patches \cite{vaswani2023attentionneed,zhou2021informer,wu2022autoformerdecompositiontransformersautocorrelation,nie2023a}. However, for battery applications directly forecasting raw terminal voltage can be inefficient because the model must learn both slow thermodynamic drift and fast pulse-driven transients. A more structured alternative is to incorporate known physical relationships analytically and learn only the residual dynamics.

Motivated by these considerations, this study develops and compares two modeling frameworks for lithium-ion battery voltage dynamics under hybrid pulse power characterization (HPPC) excitation. The first is an operator-theoretic DMDc model in which Hankel delay embeddings of voltage and current were used to identify linear state-transition and input operators directly from measurement data. The second is a physics-guided transformer-based sequence model in which terminal voltage is decomposed into an analytically computed open-circuit-voltage (OCV) component and a learned dynamic residual, and a modified PatchTST architecture is used to forecast this residual while incorporating the known future current sequence. The DMDc model is identified from healthy-state HPPC data and then applied to later-cycle datasets to assess robustness under degradation, while the transformer model is trained and evaluated using a file-level partition of the available dataset.

The objective of this work is to assess not only predictive accuracy, but also how model structure, inductive bias, and data requirements affect performance in battery voltage forecasting. In particular, the study examines whether delay-embedded DMDc can reproduce pulse-rest voltage dynamics in an interpretable state-space form, how a physics-guided transformer performs on the same class of HPPC data, and how the two approaches compare on degraded-cell voltage trajectories.

\section{Background}

This section outlines the mathematical formulation of the Hankel matrix, Dynamic Mode Decomposition with control (DMDc) algorithms and Transformer-Based Sequence Models, and how these algorithms have been applied in the context of modeling of lithium-ion battery dynamics.

\subsection{Hankel Matrix}
A hankel matrix is a special type of matrix whose skew-diagonal elements are constant.In our work, a sequence of discrete voltage signal responses $[v(1), v(2), \dots, v(N)]$,obtained from the Hybrid Pulse Power Characterization (HPPC) test, were used to form the Hankel matrix:
\begin{equation}\label{eq:Hankel}
\mathbf{H} =
\begin{bmatrix}
v(1)   & \cdots & v(n) \\
\vdots & \ddots & \vdots \\
v(m)   & \cdots & v(N)
\end{bmatrix}
\end{equation}

where $1 < m < N$, $n = N - m + 1$, and $H \in \mathbb{R}^{m \times n}$, with $m$ denoting the embedding dimension and $n$ denoting number of columns. By choosing an appropriate embedding dimension $m$ (and delay $\tau$), a block Hankel (time-delay) matrix built from $m$ consecutive delayed samples not only encodes temporal dependencies in the data, but also avoids underfitting or overfitting the system.

\subsection{Dynamic Mode Decomposition with Control}

In this study, we formed snapshots from the voltage time series data by constructing a Hankel matrix-based time-delay embedding using Equation \ref{eq:Hankel}, where each column stacks $m$ delayed samples from the hybrid pulse power characterization (HPPC) dataset and then apply DMDc to the resulting snapshot matrices. The resulting snapshot matrices are denoted by $\mathbf{X}$ and $\mathbf{X}'$: \\

\begin{equation}\label{eq:x_mat}
\mathbf{X} =
\begin{bmatrix}
v(t_{1}) & \cdots & v(t_{n-1}) \\
\vdots   & \ddots & \vdots   \\
v(t_{m}) & \cdots & v(t_{N-1})
\end{bmatrix},
\;
\mathbf{X}' =
\begin{bmatrix}
v(t_{2})   & \cdots & v(t_{n}) \\
\vdots     & \ddots & \vdots     \\
v(t_{m+1}) & \cdots & v(t_{N})
\end{bmatrix}
\end{equation}

The DMDc method aims to identify the best-fit linear operators $\mathbf{A}$ and $\mathbf{B}$ that approximately satisfy the following dynamical relationship based on the measurement data:

\begin{equation}
    \mathbf{x}_{k+1} = \mathbf{A}\mathbf{x}_{k} + \mathbf{B}\mathbf{u}_{k}
    \label{eq:model_dmdc}
\end{equation}

In addition to the snapshot matrices $\mathbf{X}$ and $\mathbf{X}'$ (Equation \ref{eq:x_mat}), a Hankel matrix representing the actuation or input history is also be constructed as follows\\

\begin{equation}\label{eq:U_mat}
\mathbf{U} =
\begin{bmatrix}
u(t_{1}) & \cdots & u(t_{n}) \\
\vdots   & \ddots & \vdots   \\
u(t_{m}) & \cdots & u(t_{N})
\end{bmatrix}
\end{equation}

The dynamics in \eqref{eq:model_dmdc} can be expressed in terms of data matrices as
\begin{equation}\label{eq:dmdc_data}
    \mathbf{X}' \approx \mathbf{A}\,\mathbf{X} + \mathbf{B}\,\mathbf{U}
\end{equation}
where \(\mathbf{X}=[\,\mathbf{x}_1,\ldots,\mathbf{x}_{N}\,]\) and \(\mathbf{X}'=[\,\mathbf{x}_2,\ldots,\mathbf{x}_{N+1}\,]\) collect measurement snapshots, and
\(\mathbf{U}=[\,\mathbf{u}_1,\ldots,\mathbf{u}_{n}\,]\) collects the corresponding input snapshots.
This formulation is advantageous for experimental studies because it requires only measured outputs and known inputs to estimate the intrinsic operator \(\mathbf{A}\) and the input map \(\mathbf{B}\).

Stacking \(\mathbf{X}\) and \(\mathbf{U}\) gives
\begin{equation}\label{eq:stacked}
    \boldsymbol{\Omega} \;\approx\;
    \begin{bmatrix}
        \mathbf{X}\\[2pt]
        \mathbf{U}
    \end{bmatrix},
    \qquad
    \mathbf{G} \;\approx\; \begin{bmatrix}\mathbf{A} & \mathbf{B}\end{bmatrix},
    \qquad
    \mathbf{X}' \approx \mathbf{G}\,\boldsymbol{\Omega}
\end{equation}
The least-squares estimate is therefore
\begin{equation}\label{eq:ls_solution}
    \mathbf{G} \;\approx\; \mathbf{X}'\,\boldsymbol{\Omega}^{\dagger}
\end{equation}
where \(^{\dagger}\) denotes the Moore--Penrose pseudoinverse.

Let the rank-\(r\) SVD of \(\boldsymbol{\Omega}\) be
\begin{equation}\label{eq:svd_omega}
    \boldsymbol{\Omega} \;\approx\; \tilde{\mathbf{U}}_r\,\tilde{\boldsymbol{\Sigma}}_r\,\tilde{\mathbf{V}}_r^{*},
    \qquad
    \boldsymbol{\Omega}^{\dagger} \;\approx\; \tilde{\mathbf{V}}_r\,\tilde{\boldsymbol{\Sigma}}_r^{-1}\,\tilde{\mathbf{U}}_r^{*}
\end{equation}
Here \(\tilde{\boldsymbol{\Sigma}}_{r}\) holds the \(r\) dominant singular values. By performing truncation, one can denoise the data and avoid inverting tiny singular values. By partitioning the left singular vectors conformably with the stacked rows, the below representation can be obtained

\begin{equation}\label{eq:U_partition}
    \tilde{\mathbf{U}}_r \;=\;
    \begin{bmatrix}
        \tilde{\mathbf{U}}_{x} & \tilde{\mathbf{U}}_{u}
    \end{bmatrix}^\top
\end{equation}
where \(\tilde{\mathbf{U}}_{x}\) spans the output (measurement) subspace and \(\tilde{\mathbf{U}}_{u}\) spans the input subspace.

\begin{figure}[]
    \centering
   
        \centering
        \includegraphics[width=\linewidth]{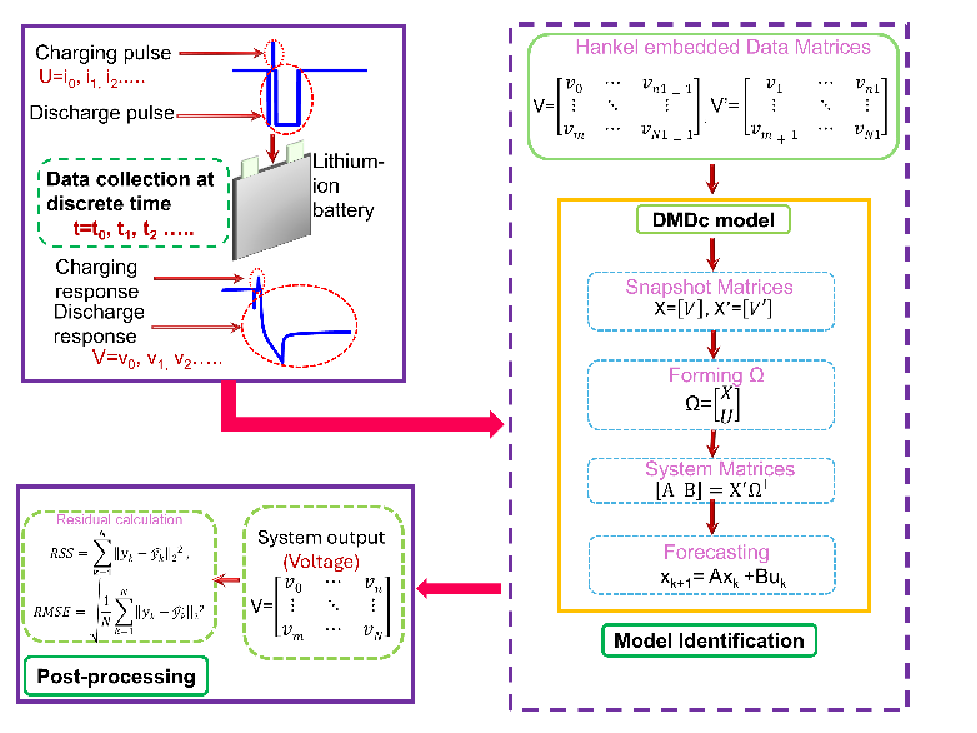}

    \caption{Schematic of the DMDc modeling framework}
    \label{fig:all cycle}
\end{figure}

Once \(\mathbf{A}\) and \(\mathbf{B}\) are obtained, the one-step ahead prediction of the dynamical system can be approximated by Equation \ref{eq:model_dmdc}. The schematic representation of the modeling framework is illustrated in Figure ~\ref{fig:all cycle}.

\subsection{Transformer-Based Sequence Models for Battery Voltage Forecasting}

Transformer-based architectures are well suited for long-horizon forecasting because self-attention can capture long-range temporal dependencies efficiently. Among recent approaches, PatchTST is particularly attractive for low-dimensional time series because patching reduces the effective sequence length from $L$ to $L/P$ while preserving local temporal structure \cite{nie2023a}. Motivated by this, the present work adopts a physics-guided PatchTST framework for battery voltage forecasting. Rather than predicting raw terminal voltage directly, the measured voltage is decomposed into an analytically computed open-circuit-voltage (OCV) component and a learned dynamic residual. The model is further adapted to the HPPC setting by introducing a future-current fusion pathway that explicitly incorporates the prescribed future current profile available at inference time.

\subsubsection{Dataset and Data Partition}

All experiments were conducted using terminal voltage and current measurements acquired from a commercial lithium-ion cell subjected to the HPPC protocol. Each file contains 20 pulse-rest cycles spanning approximately the full usable state-of-charge range of the cell, with a nominal sampling interval of about 5\,s. The available corpus comprises files up to the 160th cycle: files up to the 100th cycle were used for primary training and validation, while the remaining files were reserved for testing. To prevent temporal and file-level leakage, the train--validation split was performed at the file level. One cycle was randomly selected from the training set for validation, and the remaining cycles were used for parameter optimization. The test files remained completely unseen until final evaluation.

\begin{table}[htbp]
\centering
\caption{Summary of the HPPC dataset and file-level partition used in this study.}
\label{tab:dataset_summary}
\begin{tabular}{>{\raggedright\arraybackslash}p{0.45\linewidth} >{\centering\arraybackslash}p{0.25\linewidth}}
\toprule
\textbf{Property} & \textbf{Value} \\
\midrule
Protocol & HPPC, 20 cyc/file \\
Training split & 3 files \\
Validation split & 1 file \\
Held-out test split & 3 files \\
Sampling interval & $\approx 5$\,s \\
Voltage range & $\approx 3.1$--$4.2$\,V \\
\bottomrule
\end{tabular}
\end{table}

\subsubsection{physics-guided Voltage Decomposition}

The terminal voltage of a lithium-ion cell under load is expressed as the sum of a slowly varying equilibrium component and a fast dynamic residual:
\begin{equation}
    V(t) = V_{\mathrm{OCV}}\!\left(\mathrm{SOC}(t)\right) + V_{\mathrm{dyn}}(t)
    \label{eq:voltage_decomposition}
\end{equation}
where $V_{\mathrm{OCV}}$ denotes the open-circuit voltage as a function of state of charge, and $V_{\mathrm{dyn}}(t)$ represents the aggregate contribution of ohmic drop, charge-transfer overpotential, and diffusion-related polarization. This decomposition is central to the proposed framework: the slow thermodynamic component is computed analytically, whereas the data-driven model is tasked only with forecasting the fast residual dynamics.

This reformulation provides two practical advantages. First, it removes the dominant low-frequency drift from the forecasting target, thereby converting a strongly non-stationary problem into a more nearly stationary one. Second, it reduces the dynamic range of the regression target, which improves numerical conditioning and allows model capacity to focus on transient electrochemical behavior rather than relearning the OCV--SOC relationship from data. In the present dataset, the extracted dynamic residual is approximately zero-centered and typically remains within about $\pm 0.2$\,V.

\subsubsection{Pseudo-SOC Estimation and OCV Modeling}
\label{sec:ocv_model}

Because absolute cell capacity is not required to index the OCV trajectory, the implementation uses a relative pseudo-state-of-charge obtained through Coulomb counting followed by per-file normalization. Let $\Delta t_k$ denote the elapsed time between samples in seconds and $I(t_k)$ the measured current. A cumulative charge proxy is first computed as
\begin{equation}
    q(t_k) = - \sum_{j=1}^{k} \frac{I(t_j)\,\Delta t_j}{3600}
    \label{eq:raw_soc}
\end{equation}
where the negative sign is consistent with the discharge-current sign convention adopted in the data. The pseudo-SOC is then normalized to the unit interval:
\begin{equation}
    \mathrm{SOC}(t_k) =
    \frac{q(t_k) - \min\limits_j q(t_j)}
         {\max\limits_j q(t_j) - \min\limits_j q(t_j)}
    \label{eq:pseudo_soc}
\end{equation}
This normalization preserves the monotonic ordering of charge depletion within each file and is sufficient for mapping the trajectory onto an empirical OCV lookup curve.

The OCV model is estimated directly from HPPC rest segments. For each training file, contiguous samples satisfying $|I(t)| \leq 0.02$\,A are identified, and only segments with duration of at least 60\,s are retained. From each valid rest segment, the voltage at the final sample is paired with the corresponding pseudo-SOC, since this endpoint provides the closest available approximation to equilibrium voltage. These samples are aggregated across the training files and projected onto a uniform 128-point SOC grid in $[0,1]$. The OCV at each grid location is computed as the mean voltage of samples within a half-bin neighborhood; when no samples are available, nearest-neighbor interpolation is used as a fallback. The resulting piecewise-linear curve is stored as a compact lookup table and evaluated during both training and inference using one-dimensional interpolation.

For any time series, the dynamic voltage is then defined as
\begin{equation}
    V_{\mathrm{dyn}}(t) = V_{\mathrm{meas}}(t) - V_{\mathrm{OCV}}\!\left(\mathrm{SOC}(t)\right)
    \label{eq:vdyn}
\end{equation}
which serves as the prediction target for the neural forecasting model.

\subsubsection{physics-guided PatchTST Architecture}
\label{sec:model}

\begin{figure}[]
    \centering
    \includegraphics[width= 0.9 \linewidth]{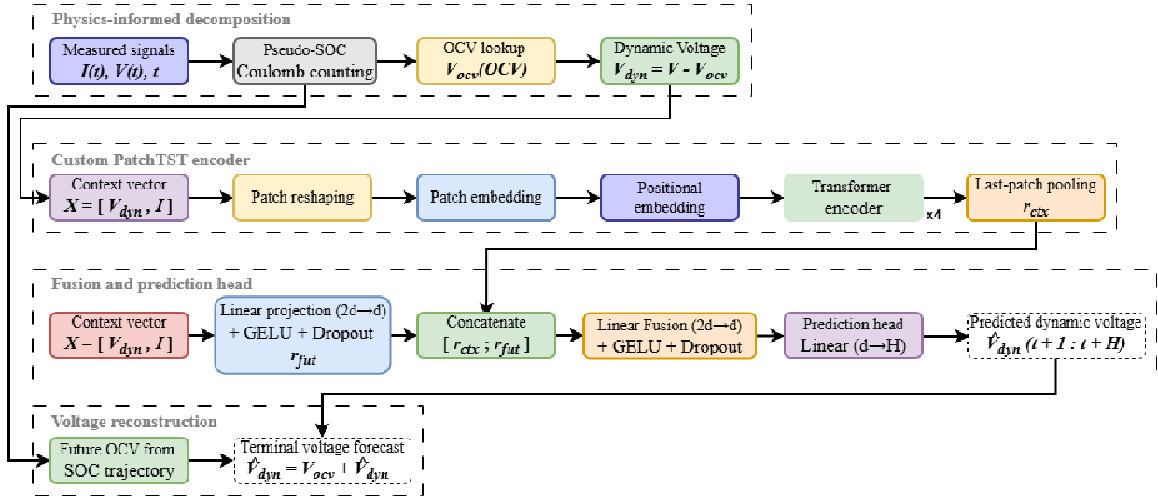}
\caption{Proposed physics-guided PatchTST for multi-step terminal-voltage forecasting under HPPC profiles}
    \label{fig:system_framework}
\end{figure}

The neural forecasting component is a modified PatchTST model designed for direct multi-step prediction of $V_{\mathrm{dyn}}$. In contrast to the original PatchTST formulation~\cite{nie2023a}, the present implementation processes a two-channel context sequence, employs global z-score normalization rather than reversible instance normalization, pools the final patch representation instead of flattening all encoder outputs, and introduces a dedicated future-covariate branch for the known HPPC current profile. The overall physics-guided PatchTST forecasting framework is illustrated in Figure~\ref{fig:system_framework}.

For batch size $B$, context length $L$, and prediction horizon $H$, the model input consists of (i) a context tensor
\begin{equation}
    \mathbf{X} \in \mathbb{R}^{B \times L \times 2},
    \qquad
    \mathbf{X}[b,t,:] = \left[\tilde{V}_{\mathrm{dyn}}(t),\,\tilde{I}(t)\right]
    \label{eq:context_tensor}
\end{equation}
and (ii) a future-current vector
\begin{equation}
    \mathbf{c}_{\mathrm{fut}} \in \mathbb{R}^{B \times H}
    \label{eq:future_current}
\end{equation}
where the tildes denote normalization using statistics computed only from the training files. The context is partitioned into $N=L/P$ non-overlapping patches of length $P$, flattened across the two input channels, and projected into the model space:
\begin{equation}
    \begin{aligned}
        \mathbf{X}_{\mathrm{patch}} &\in \mathbb{R}^{B \times N \times 2P}, \\
        \mathbf{Z}^{(0)} = \mathbf{X}_{\mathrm{patch}} \mathbf{W}_{\mathrm{emb}} + \mathbf{b}_{\mathrm{emb}}
        &\in \mathbb{R}^{B \times N \times d_{\mathrm{model}}}
    \end{aligned}
    \label{eq:patch_embedding}
\end{equation}
Learned positional embeddings are added to the patch embeddings, followed by dropout.

The embedded patch sequence is passed through $N_L$ pre-normalized Transformer encoder layers. Each layer applies multi-head self-attention and a position-wise feed-forward network with residual connections:
\begin{equation}
\label{eq:transformer_block}
\begin{aligned}
(\mathbf{Z}')^{(\ell)} &= \mathbf{Z}^{(\ell-1)} + \mathrm{MHSA}\!\left(\mathrm{LN}\!\left(\mathbf{Z}^{(\ell-1)}\right)\right), \\
\mathbf{Z}^{(\ell)} &= (\mathbf{Z}')^{(\ell)} + \mathrm{FFN}\!\left(\mathrm{LN}\!\left((\mathbf{Z}')^{(\ell)}\right)\right).
\end{aligned}
\end{equation}
The self-attention operator is defined as
\begin{equation}
    \begin{aligned}
        \mathrm{Attention}(\mathbf{Q},\mathbf{K},\mathbf{V})
        &=
        \mathrm{softmax}\!\left(
        \frac{\mathbf{Q}\mathbf{K}^{\top}}{\sqrt{d_k}}
        \right)\mathbf{V}, \\
        d_k &= \frac{d_{\mathrm{model}}}{h}
    \end{aligned}
    \label{eq:attention}
\end{equation}
The feed-forward block uses GELU activation:
\begin{equation}
    \mathrm{FFN}(\mathbf{x}) =
    \mathrm{GELU}\!\left(\mathbf{x}\mathbf{W}_1 + \mathbf{b}_1\right)\mathbf{W}_2 + \mathbf{b}_2
    \label{eq:ffn}
\end{equation}
After the final encoder layer, LayerNorm is applied and the representation of the last patch is extracted as a context summary vector $\mathbf{r}_{\mathrm{ctx}} \in \mathbb{R}^{B \times d_{\mathrm{model}}}$

A key architectural extension is the future-current fusion pathway. Since the HPPC current waveform is prescribed by the experimental protocol, the next $H$ current values are known at inference time and can be incorporated explicitly. The normalized future-current vector is projected into the latent space as
\begin{equation}
    \mathbf{r}_{\mathrm{fut}}
    =
    \mathrm{GELU}\!\left(
    \mathbf{c}_{\mathrm{fut}}\mathbf{W}_{\mathrm{fut}} + \mathbf{b}_{\mathrm{fut}}
    \right)
    \in \mathbb{R}^{B \times d_{\mathrm{model}}}
    \label{eq:future_projection}
\end{equation}
after which the context representation and future-current embedding are concatenated and fused through a learned linear transformation:
\begin{equation}
    \mathbf{r}_{\mathrm{fused}}
    =
    \mathrm{GELU}\!\left(
    [\mathbf{r}_{\mathrm{ctx}};\mathbf{r}_{\mathrm{fut}}]\mathbf{W}_{\mathrm{fuse}} + \mathbf{b}_{\mathrm{fuse}}
    \right)
    \in \mathbb{R}^{B \times d_{\mathrm{model}}}
    \label{eq:fusion}
\end{equation}
A final linear head produces the $H$-step forecast of the normalized dynamic voltage:
\begin{equation}
    \hat{\mathbf{v}}_{\mathrm{dyn}}
    =
    \mathbf{r}_{\mathrm{fused}}\mathbf{W}_{\mathrm{head}} + \mathbf{b}_{\mathrm{head}}
    \in \mathbb{R}^{B \times H}
    \label{eq:prediction_head}
\end{equation}

For the reported model, the architecture uses $L=1024$, $H=64$, $P=16$, $d_{\mathrm{model}}=256$, $N_L=4$, $h=8$, and $d_{\mathrm{ff}}=1024$, resulting in approximately $3.35 \times 10^6$ trainable parameters.

\subsubsection{Training Procedure}

Training is performed on overlapping sliding windows extracted from the dynamic-voltage sequences. For each file, the fitted OCV model is first used to compute $V_{\mathrm{dyn}}(t)$ at every sample. A window of total length $L+H$ is then moved across the sequence with stride 1. Each training sample consists of (i) a context tensor formed from the previous $L$ normalized values of $V_{\mathrm{dyn}}$ and current, (ii) the next $H$ normalized current values, and (iii) the next $H$ normalized dynamic-voltage targets. This windowing strategy provides direct multi-step supervision and avoids the instability associated with teacher-forced one-step decoding during training.

The model is trained by minimizing the mean squared error on normalized dynamic voltage:
\begin{equation}
    \mathcal{L}
    =
    \frac{1}{BH}
    \sum_{b=1}^{B}
    \sum_{k=1}^{H}
    \left(
    \hat{v}_{\mathrm{dyn}}^{(b,k)} - v_{\mathrm{dyn}}^{(b,k)}
    \right)^2
    \label{eq:mse_loss}
\end{equation}
Optimization uses AdamW~\cite{loshchilov2019decoupledweightdecayregularization} with learning rate $3.0\times10^{-4}$, weight decay $1.0\times10^{-5}$, cosine annealing to a minimum learning rate of $3.0\times10^{-6}$~\cite{loshchilov2017sgdrstochasticgradientdescent}, and gradient clipping with maximum norm 1.0. Dropout with rate 0.1 is applied in the embedding, attention, feed-forward, future-projection, and fusion layers. Early stopping with patience 15 epochs is used on validation loss. During training, the optimal model checkpoint was obtained at epoch 37, after which no further improvement in validation loss was observed; training was subsequently terminated at epoch 52 upon exhausting the early stopping patience of 15 epochs.

\begin{table}[htbp]
\centering
\caption{Hyperparameters of the reported physics-guided PatchTST model.}
\label{tab:hyperparameters}
\begin{tabular}{lll}
\toprule
\textbf{Category} & \textbf{Hyperparameter} & \textbf{Value} \\
\midrule
Architecture & Context length $L$ & 1024 \\
Architecture & Prediction horizon $H$ & 64 \\
Architecture & Patch length $P$ & 16 \\
Architecture & Number of patches $N$ & 64 \\
Architecture & Model dimension $d_{\mathrm{model}}$ & 256 \\
Architecture & Attention heads $h$ & 8 \\
Architecture & Encoder layers $N_L$ & 4 \\
Architecture & Feed-forward dimension $d_{\mathrm{ff}}$ & 1024 \\
Architecture & Activation & GELU \\
Architecture & Input channels & 2 ($V_{\mathrm{dyn}}$, current) \\
Optimization & Optimizer & AdamW \\
Optimization & Gradient clipping & 1.0 \\
Optimization & Batch size & 32 \\
Regularization & Dropout & 0.1 \\
Model size & Trainable parameters & 3,348,800 \\
\bottomrule
\end{tabular}
\end{table}

\section{Experimental set-up}

A 30 Ah lithium-ion battery was employed to generate hybrid pulse power characterization (HPPC) test data for the purpose of modeling and analyzing the battery's dynamic behavior. The battery operated within a voltage range of 2.5 V to 4.2 V. Initially, the battery was fully charged using a two-step charging protocol consisting of: (1) constant current (CC) charging, followed by (2) constant voltage (CV) charging. Subsequently, the battery was allowed to rest for one hour (Step 3). A 10 A constant current discharge pulse was then applied for 10 seconds (Step 4), after which the battery rested for 3 minutes (Step 5). This was followed by a 5 A constant current charge pulse applied for 20 seconds (Step 6), and another rest period of 2 minutes (Step 7). Next, a 10 A constant current discharge was carried out for 18 minutes, corresponding to a 0.1 Ah capacity reduction, terminating when the battery voltage reached 3.95 V. After this, the battery was rested for 1 hour, during which the voltage gradually recovered to approximately 4.07 V (Step 8). Steps 3 through 8 were repeated 10 times to simulate various state of charge (SoC). Following the completion of the HPPC test on a fresh (healthy) battery, the cell was subjected to continuous cycling to induce capacity degradation. After every 20 charge-discharge cycles, an HPPC test (using the same procedure as described above) was conducted to assess changes in the battery's voltage response, thereby characterizing the effects of degradation over time.

\section{Results and Discussions}

In this section, we report comparison of simulation results obtained from dynamic mode decomposition with control (DMDc) with transformer-based sequence learning architecture. We describe model identification (rank/embedding selection and least-squares fitting) and present predictions of impulse charge-discharge behavior, followed by an analysis of the identified dynamics. %(eigenvalues/modes and input–output effects).

For experimental validation, the model-predicted charge-discharge profiles are compared with the measured profiles under identical input sequences. The predictive performance is evaluated using two commonly used error metrics: the residual sum of squares (RSS) and the root mean square error (RMSE). The RSS is defined as

\begin{equation}\label{RSS}
\mathrm{RSS} = \sum_{k=1}^{N} \left\| \mathbf{y}_k - \hat{\mathbf{y}}_k \right\|_2^{2}
\end{equation}

And the root mean square error (RMSE) is then computed as

\begin{equation}\label{RMSE}
\mathrm{RMSE} = \sqrt{\frac{1}{N} \sum_{k=1}^{N} \left\| \mathbf{y}_k - \hat{\mathbf{y}}_k \right\|_2^{2}}
\end{equation}

where $\mathbf{y}_k$ and $\hat{\mathbf{y}}_k$ denote the measured and predicted outputs at time step $k$, respectively, and $N$ represents the total number of observations. For scalar outputs, the squared $\ell_2$-norm reduces to the squared difference between the measured and predicted values.
\subsection{Dynamic Mode Decomposition with Control}
\subsubsection{Hankel embedded data matrices}
In our study, only a single output (voltage) is available, and the true state is unobserved. So to enrich the observable coordinates, we constructed block Hankel (time–delay) embedded snapshot matrices \(\mathbf{X}=\mathcal{H}_m(y)\), according to the Equation \ref{eq:Hankel} mentioned earlier, which lifted the temporal structure into an \(m\)-dimensional Koopman-observable space.

\begin{figure}[t]
    \centering

  \subfigure{
        \includegraphics[width=0.49\linewidth]{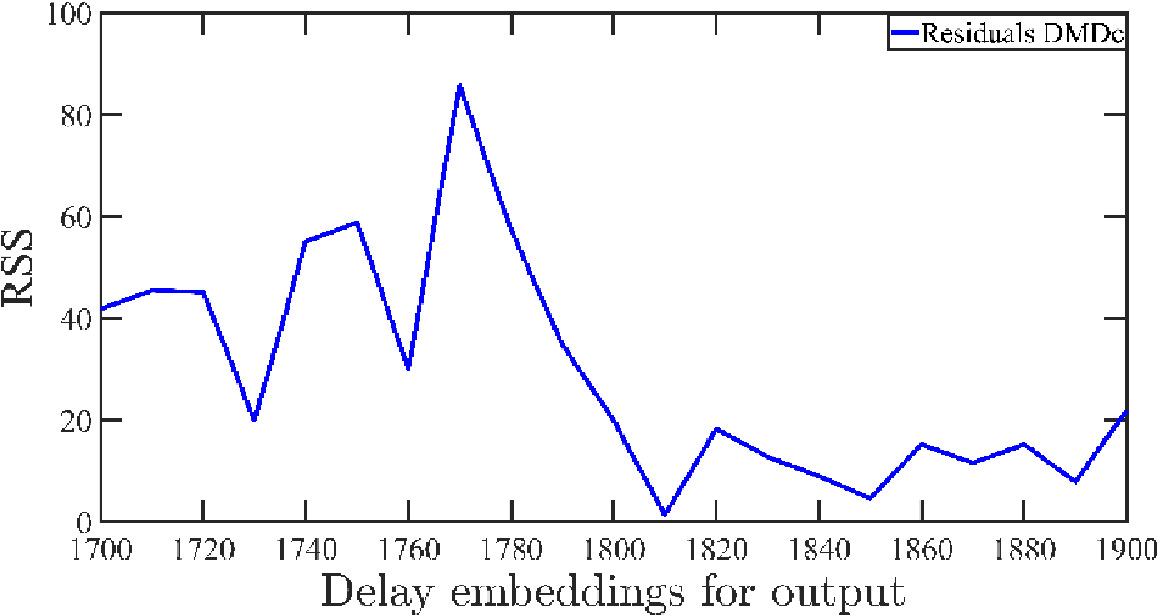}
        \label{fig:output}
}
    \hfill
 \subfigure{
        \includegraphics[width=0.46\linewidth]{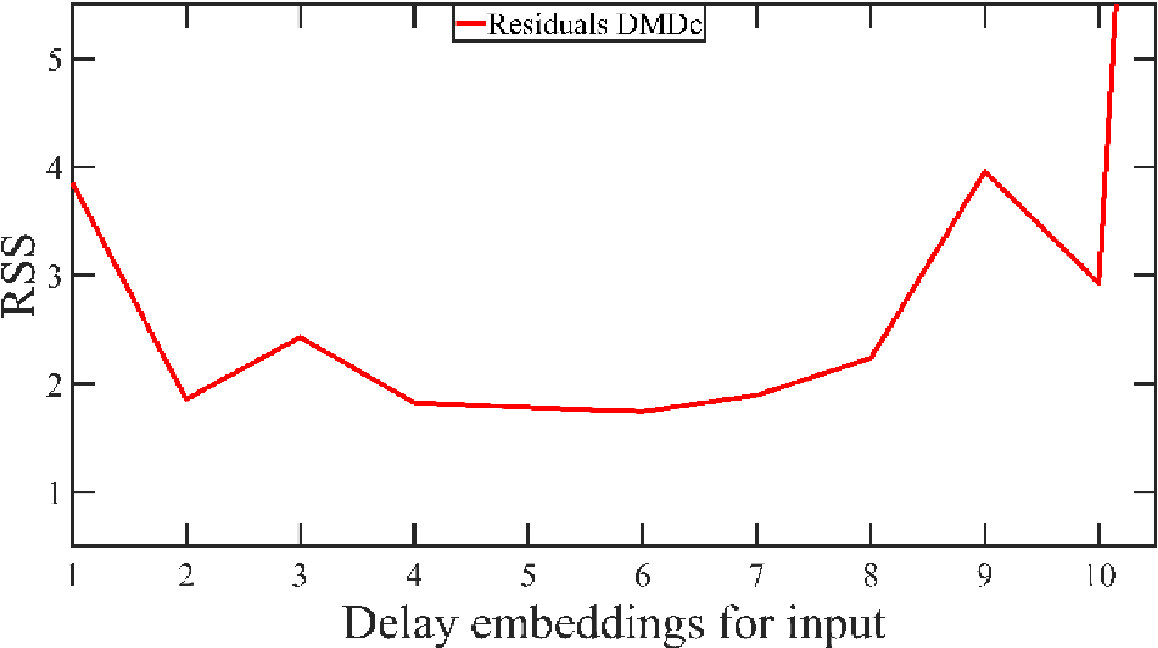}
        \label{fig:input}
 
}
    \caption{Comparison of RSS for different delay embeddings (a) Voltage (output) (b) current (input) DMDc models}
    \label{fig:RSS1}
\end{figure}

We swept the embedding dimension \(m\in[1700,1900]\) and computed the residual sum of squares (RSS) for each case. The lowest residual obtained for DMD with control (DMDc), i.e., \(\mathbf{X}' \approx \mathbf{A}\mathbf{X}+\mathbf{B}\mathbf{U}\) was 3.85. To further enhance performance, we also delay-embedded the input to form \(\mathbf{U}=\mathcal{H}_{m}(u)\) and tested \(m=1,\dots,12\); a six-delay input embedding yielded the lowest residual of 1.74, as shown in Figure \ref{fig:RSS1}(b).

\subsubsection{Performance of the Model at Degraded States}

For model training, 60\% of the voltage and current data from healthy state cycle were used. The system transition matrix $\mathbf{A}$ and the input transition matrix $\mathbf{B}$ for the DMDc model, were computed using the equations described earlier. Once these matrices were estimated and the appropriate initial conditions specified, $\mathbf{A}$ and $\mathbf{B}$ were applied iteratively to the augmented state vector to approximate the system dynamics over time. This linearized model was then employed for multi-step, look-ahead predictions, with results compared against experimental measurements.

\begin{figure}[t]
    \centering

    \subfigure{
        \includegraphics[width=0.48\linewidth]{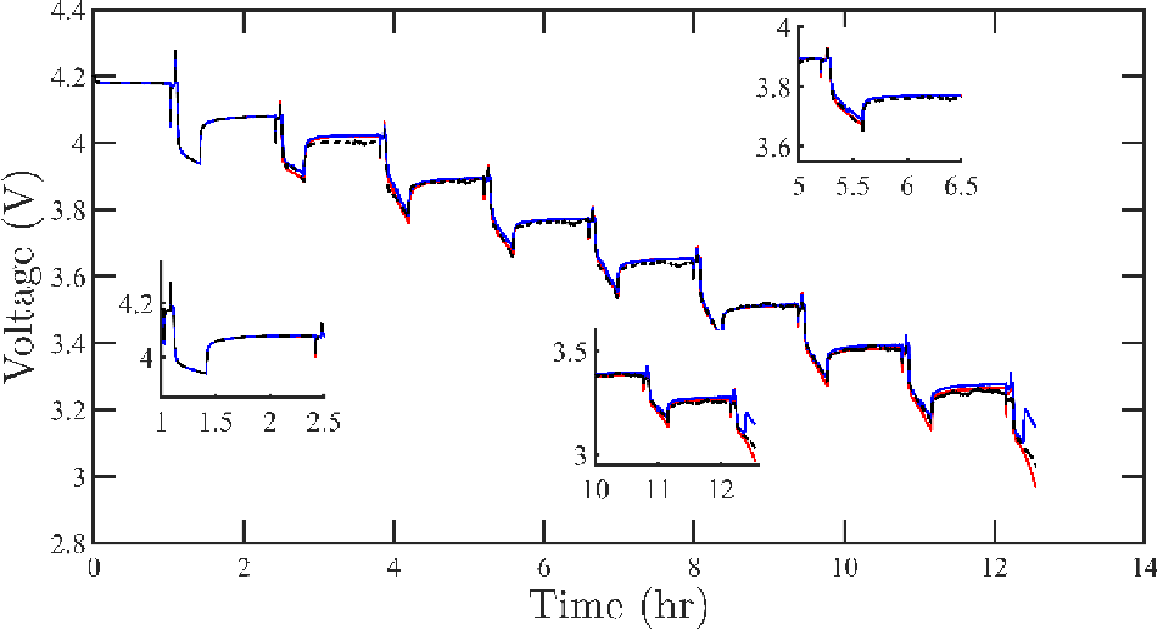}
        \label{fig:cycle_120}
    }
    \hfill
    \subfigure{
        \includegraphics[width=0.48\linewidth]{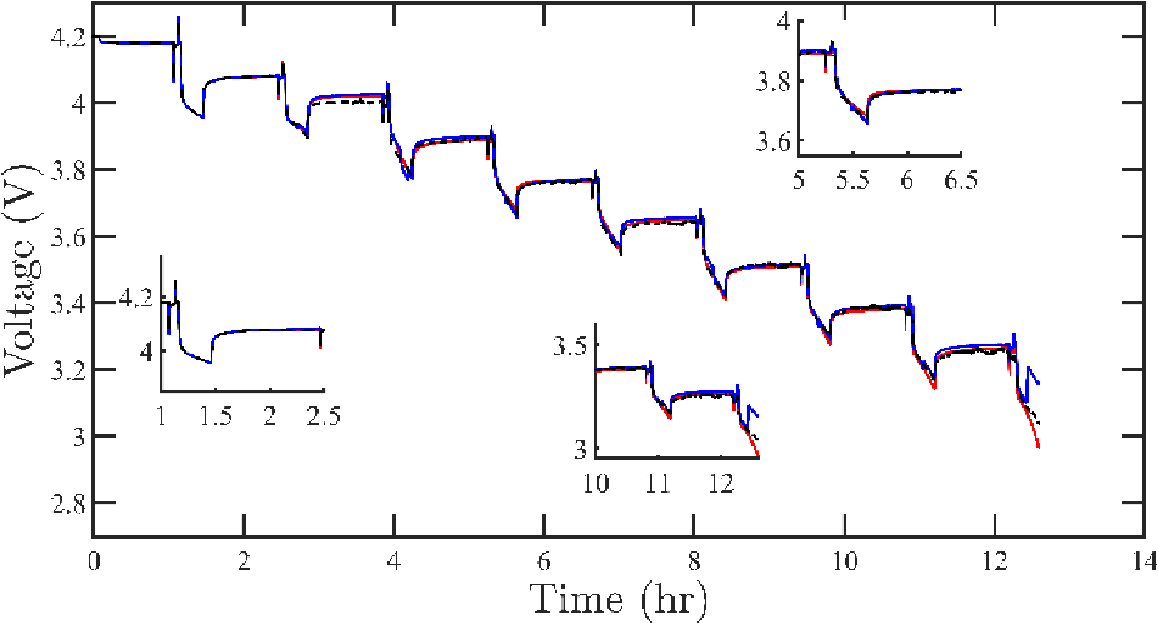}
        \label{fig:cycle_140}
    }
    \hfill
    \subfigure{
        \includegraphics[width=0.48\linewidth]{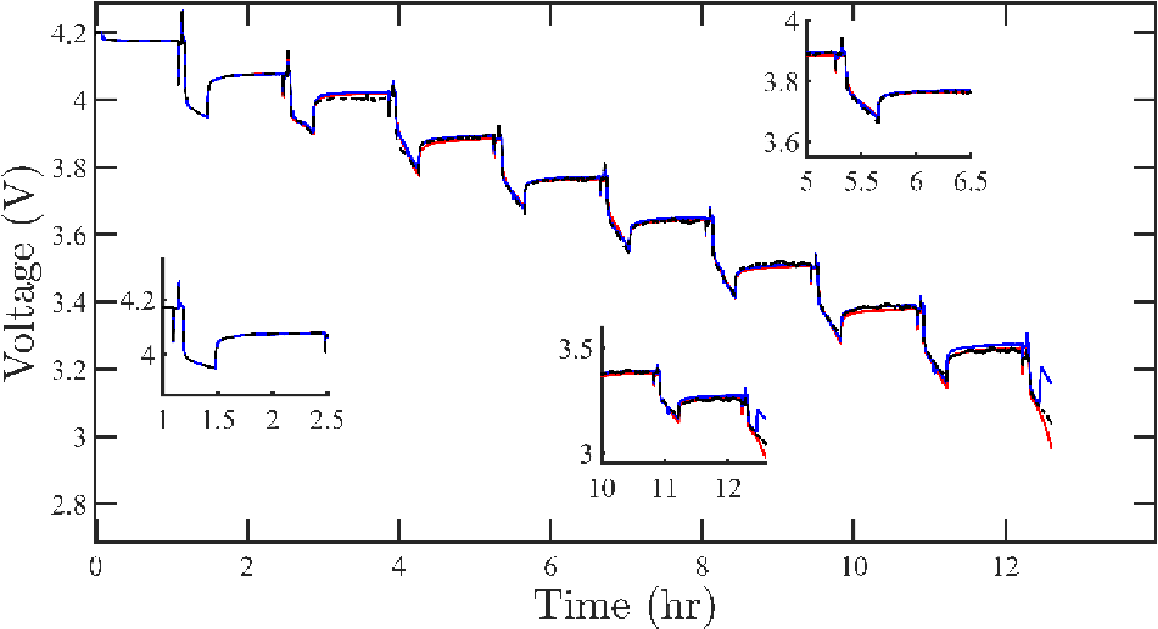}
        \label{fig:cycle_160}
    }

    \vspace{0.2cm}

    \includegraphics[width=0.33\linewidth]{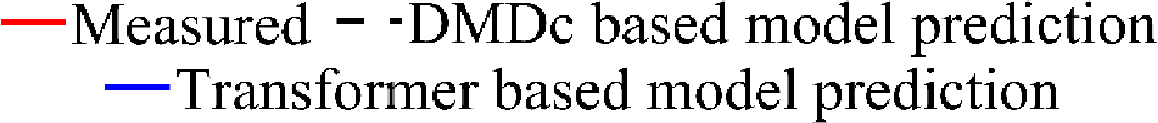}

    \caption{Comparison of experimental measurements and model predictions for (a) the 120th cycle, (b) the 140th cycle, and (c) the 160th cycle.}
    \label{fig:degraded_model}
\end{figure}

Figure~\ref{fig:degraded_model}(a) presents the complete voltage trajectory obtained from the HPPC test alongside the corresponding predictions of the identified DMDc model for the 120th cycle. At each time step, the model advances the system state using its previously predicted values. Despite this sequential propagation, the predicted voltage closely follows the experimental measurements throughout the full time horizon.If looked at the zoomed segments (Figure \ref{fig:degraded_model}(a)), in the initial stage (up to \(\sim 2.5\,\mathrm{h}\)), predictions from the model matched the measurements perfectly. As the forecast horizon increased, discrepancies started to be visible, however, model maintained accurate pulse-response predictions with very little negligible deviation in the slow relaxation segment. Even in the late stage (after \(\sim 10\,\mathrm{h}\)), it demonstrated strong long-horizon performance, achieving RSS of 1.1381 and RMSE of 0.012 even for far-future predictions.

\begin{figure}[tbp]
    \centering
    \includegraphics[width=0.75\linewidth]{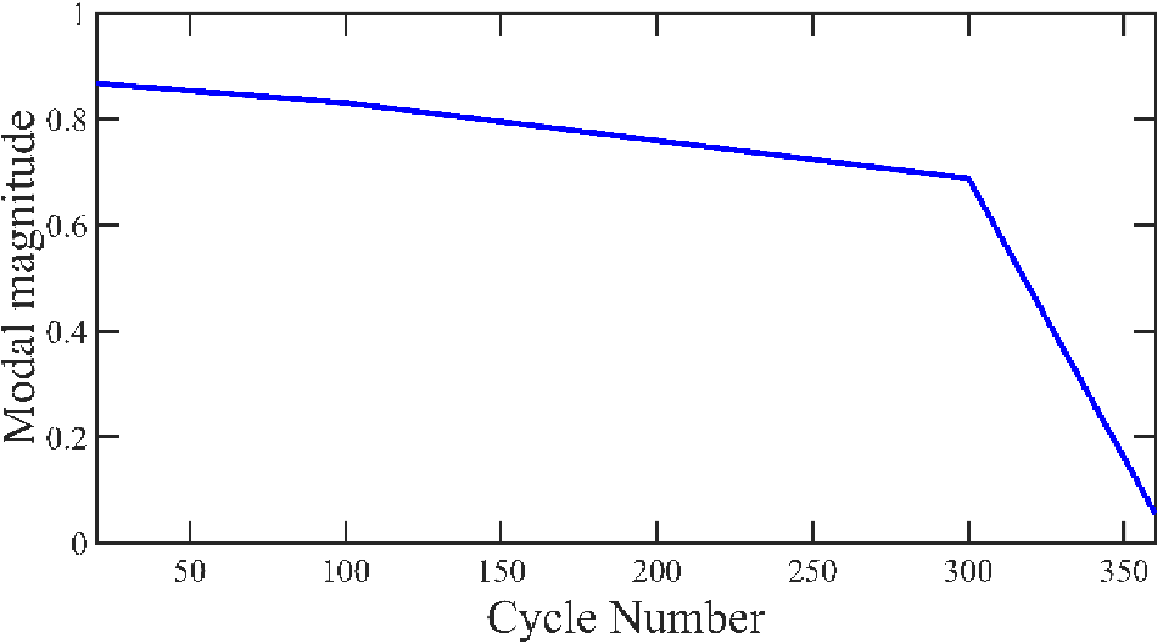}
    \caption{Change of modal magnitude with cycles}
    \label{fig:aging}
\end{figure}

 Figures~\ref{fig:degraded_model}(b) and \ref{fig:degraded_model}(c) further illustrate the model’s robustness at more advanced degradation stages, specifically for the 140th and 160th cycles, respectively. As summarized in Table~\ref{tab:model_comparison}, the model maintains a high level of accuracy, as well as effectively capturing both the sharp transient impulses and the subsequent relaxation behavior of the discharge profile, even at degraded states.
Moreover,  the magnitudes of the identified dynamic modes, quantified by the 
modal magnitude $r = |\lambda|$, provide direct insight into the dissipative behavior of the system across aging stages. As shown in Figure \ref{fig:aging}, progressive battery aging is associated with a systematic decrease in modal magnitudes, reflecting increasingly dissipative dynamics over the cycling history.

\subsection{Transformer-Based Sequence Models}

Model evaluation is performed on the reconstructed terminal voltage rather than on the residual alone. For each held-out test file, pseudo-SOC and OCV are computed over the entire sequence using the analytical physics layer. The first 15\% of the file is used as an observed context region, while the remaining 85\% constitutes the evaluation region. Prediction then proceeds auto-regressively in chunks of length $H$. At each rollout step, the model receives the most recent context window of length $L$, together with the known future current over the next $H$ samples, and returns a direct $H$-step forecast of $V_{\mathrm{dyn}}$. These predicted residuals are written back into a running buffer and then added to the analytical OCV trajectory to reconstruct terminal voltage.

This chunked rollout strategy improves computational efficiency because each forward pass yields multiple future samples, yet it still preserves the realistic deployment condition that, beyond the initial context, subsequent predictions depend on the model's own previous outputs. Let $s$ denote the evaluation start index and let $T$ denote the sequence length. The inference procedure is summarized in Algorithm~\ref{alg:rollout}.

\begin{algorithm}[htbp]
\caption{Chunked autoregressive inference for physics-guided PatchTST}
\label{alg:rollout}
\small
\setlength{\abovedisplayskip}{0pt}
\begin{algorithmic}[1]
\setlength{\itemsep}{0pt}
\setlength{\parsep}{0pt}
\Require Current $I_{1:T}$, measured voltage $V_{1:T}$, time stamps $t_{1:T}$, fitted OCV model, trained PatchTST forecaster, context length $L$, chunk length $H$, evaluation start index $s$
\Ensure Reconstructed voltage forecast $\hat{V}_{1:T}$
\State Compute pseudo-SOC trajectory $\mathrm{SOC}_{1:T}$ using Eqs.~\eqref{eq:raw_soc}--\eqref{eq:pseudo_soc}
\State Compute $V_{\mathrm{OCV},1:T}$ by interpolating OCV--SOC lookup table
\State Compute dynamic voltage $V_{\mathrm{dyn},1:T} = V_{1:T} - V_{\mathrm{OCV},1:T}$
\State Initialize $\hat{V}_{1:s-1} \gets V_{1:s-1}$ and residual buffer $\mathbf{b} \gets V_{\mathrm{dyn},1:T}$
\State Set $t \gets s$
\While{$t \leq T$}
    \State $\ell \gets \min(H,\; T-t+1)$
    \State Extract residual history $\mathbf{b}_{t-L:t-1}$, current history $I_{t-L:t-1}$
    \State Extract known future current $I_{t:t+\ell-1}$
    \State Predict $\hat{\mathbf{v}}_{\mathrm{dyn}} \gets \mathrm{PatchTST}(\mathbf{b}_{t-L:t-1},\, I_{t-L:t-1},\, I_{t:t+\ell-1})$
    \State Update residual buffer $\mathbf{b}_{t:t+\ell-1} \gets \hat{\mathbf{v}}_{\mathrm{dyn}}$
    \State Reconstruct voltage $\hat{V}_{t:t+\ell-1} \gets V_{\mathrm{OCV},t:t+\ell-1} + \hat{\mathbf{v}}_{\mathrm{dyn}}$
    \State $t \gets t + \ell$
\EndWhile
\end{algorithmic}
\end{algorithm}

Performance is reported on the reconstructed voltage over the evaluation region only, using root mean squared error (RMSE), and residual sum of squares (RSS). Aggregate metrics are obtained by concatenating the evaluation regions of all three held-out test files before metric computation.

\begin{table}[htbp]
\centering
\caption{Comparison of prediction error metrics for DMDc and transformer-based models.}
\label{tab:model_comparison}
\begin{tabular}{lcccc}
\toprule
\multirow{2}{*}{\textbf{Cycle}} & \multicolumn{2}{c}{\textbf{DMDc Model}} & \multicolumn{2}{c}{\textbf{Transformer-Based}} \\
\cmidrule(lr){2-3} \cmidrule(lr){4-5}
 & RSS & RMSE & RSS & RMSE \\
\midrule
120 & 1.1381 & 0.0112 & 3.44 & 0.0211 \\
140 & 1.0800 & 0.0109 & 3.69 & 0.0218 \\
160 & 1.0460 & 0.0107 & 3.45 & 0.0211 \\
\bottomrule
\end{tabular}
\end{table}

The lower accuracy of the transformer-based model reported in Table~\ref{tab:model_comparison} should be interpreted in light of both the available data regime and the inductive bias of the competing modeling frameworks. Although the sliding-window strategy yields a large number of training samples, the effective diversity of the neural training set remains limited because these samples are generated from a small number of HPPC files collected under a single protocol and within a relatively narrow range of operating conditions. Under such conditions, the PatchTST model, with approximately $3.35\times10^6$ trainable parameters, is likely more difficult to train robustly than the DMDc formulation, which imposes a much stronger structural prior through linear controlled evolution in a delay-embedded state space. This prior appears to be particularly well aligned with the repeatable pulse-rest transients characteristic of HPPC data, whereas the transformer must infer these relationships directly from data and is additionally exposed to error accumulation during chunked autoregressive rollout.

At the same time, the present findings should not be interpreted as evidence that transformer-based approaches are inherently inferior for battery voltage forecasting. Their primary advantage lies in their flexibility: with substantially larger and more heterogeneous datasets spanning multiple cells, aging trajectories, temperature conditions, and operating protocols, such models may be better able to capture nonlinear dependencies that are difficult to represent through a single linear operator. Whether this potential advantage can be realized, however, depends critically on access to sufficiently diverse training data together with appropriate regularization and model selection; it is not guaranteed by model class alone.

\section{Conclusions} 

This study compared two data-driven frameworks for modeling lithium-ion battery voltage dynamics under HPPC excitation: DMDc and a physics-guided transformer-based sequence model. In the DMDc framework, Hankel delay embeddings of voltage and current were used to identify linear system matrices from healthy-state data, and the resulting model was applied to later degraded-cycle datasets. In the transformer framework, terminal voltage was decomposed into an analytically computed OCV component and a dynamic residual, and a modified PatchTST model was trained to predict the residual over a finite horizon.

Both approaches were able to reproduce the overall pulse-rest voltage behavior, but DMDc achieved lower prediction error and greater robustness under the present data regime. In particular, the DMDc model maintained accurate predictions when applied to degraded-cell data using operators identified from healthy-state HPPC measurements. The transformer-based model remained flexible and capable of capturing the same qualitative dynamics, but its performance was more limited by the size and diversity of the available training data. In addition, the identified DMDc eigensvalues exhibited a monotonic reduction in modal magnitudes across aging stages, suggesting potential utility for degradation characterization.

Overall, the results indicate that DMDc provides a favorable balance of accuracy, robustness, interpretability, and computational efficiency for battery voltage forecasting under HPPC excitation. More broadly, the study highlights the promise of operator-theoretic methods for adaptive battery diagnostics and prognostics, while also indicating that physics-guided transformer models may become increasingly competitive as larger and more heterogeneous datasets become available.

\bibliographystyle{unsrt}     
\bibliography{references}  

@article{Xiong2018,
  title = {Towards a smarter battery management system: A critical review on battery state of health monitoring methods},
  volume = {405},
  ISSN = {0378-7753},
  url = {http://dx.doi.org/10.1016/j.jpowsour.2018.10.019},
  DOI = {10.1016/j.jpowsour.2018.10.019},
  journal = {Journal of Power Sources},
  publisher = {Elsevier BV},
  author = {Xiong,  Rui and Li,  Linlin and Tian,  Jinpeng},
  year = {2018},
  month = nov,
  pages = {18–29}
}

@article{Wang2020,
  title = {A comprehensive review of battery modeling and state estimation approaches for advanced battery management systems},
  volume = {131},
  ISSN = {1364-0321},
  url = {http://dx.doi.org/10.1016/j.rser.2020.110015},
  DOI = {10.1016/j.rser.2020.110015},
  journal = {Renewable and Sustainable Energy Reviews},
  publisher = {Elsevier BV},
  author = {Wang,  Yujie and Tian,  Jiaqiang and Sun,  Zhendong and Wang,  Li and Xu,  Ruilong and Li,  Mince and Chen,  Zonghai},
  year = {2020},
  month = oct,
  pages = {110015}
}

@inproceedings{FARHANGDOUST2023,
  series = {shm2023},
  title = {DESIGN OF MULTIFUNCTIONAL STRUCTURAL BATTERY COMPOSITES FOR THE NEXT GENERATION OF ELECTRIC VEHICLES},
  url = {http://dx.doi.org/10.12783/shm2023/36814},
  DOI = {10.12783/shm2023/36814},
  booktitle = {Proceedings of the 14th International Workshop on Structural Health Monitoring},
  publisher = {Destech Publications,  Inc.},
  author = {Farhangdoust, Saman and Ahmed, Shabbir and Strange, Alexander and Altuntas, Umut and Duan, Chaoqun and Abdullah, Yaqoub and Li, Franklin and Wang, Serena and Chang, Fu-Kuo},
  year = {2023},
  month = sep,
  collection = {shm2023}
}

@article{quelin2023coupling,
  title={Coupling electrical parameters of a battery equivalent circuit model to electrodes dimensions},
  author={Quelin, Aur{\'e}lien and Damay, Nicolas},
  journal={Journal of Power Sources},
  volume={561},
  pages={232690},
  year={2023},
  publisher={Elsevier}
}

@article{tran2021comprehensive,
  title={A comprehensive equivalent circuit model for lithium-ion batteries, incorporating the effects of state of health, state of charge, and temperature on model parameters},
  author={Tran, Manh-Kien and Mathew, Manoj and Janhunen, Stefan and Panchal, Satyam and Raahemifar, Kaamran and Fraser, Roydon and Fowler, Michael},
  journal={Journal of Energy Storage},
  volume={43},
  pages={103252},
  year={2021},
  publisher={Elsevier}
}

@inproceedings{petri2023state,
  title={State estimation of an electrochemical lithium-ion battery model: improved observer performance by hybrid redesign},
  author={Petri, Elena and Reynaudo, Thomas and Postoyan, Romain and Astolfi, Daniele and Ne{\v{s}}i{\'c}, D and Ra{\"e}l, St{\'e}phane},
  booktitle={2023 European Control Conference (ECC)},
  pages={1--6},
  year={2023},
  organization={IEEE}
}

@article{ahmed2025autoregressive,
  title={Autoregressive model-based parameter correlation for state of charge and state of health of lithium-ion batteries using built-in piezoelectric transducer induced ultrasonic waves},
  author={Ahmed, Shabbir and Farhangdoust, Saman and Chang, Fu-Kuo},
  journal={Journal of Energy Storage},
  volume={114},
  pages={115829},
  year={2025},
  publisher={Elsevier}
}

@article{lai2018comparative,
  title={A comparative study of different equivalent circuit models for estimating state-of-charge of lithium-ion batteries},
  author={Lai, Xin and Zheng, Yuejiu and Sun, Tao},
  journal={Electrochimica Acta},
  volume={259},
  pages={566--577},
  year={2018},
  publisher={Elsevier}
}

@article{kaiser2018sparse,
  title={Sparse identification of nonlinear dynamics for model predictive control in the low-data limit},
  author={Kaiser, Eurika and Kutz, J Nathan and Brunton, Steven L},
  journal={Proceedings of the Royal Society A},
  volume={474},
  number={2219},
  pages={20180335},
  year={2018},
  publisher={The Royal Society Publishing}
}

@article{rudy2017data,
  title={Data-driven discovery of partial differential equations},
  author={Rudy, Samuel H and Brunton, Steven L and Proctor, Joshua L and Kutz, J Nathan},
  journal={Science advances},
  volume={3},
  number={4},
  pages={e1602614},
  year={2017},
  publisher={American Association for the Advancement of Science}
}

@article{piroddi2003identification,
  title={An identification algorithm for polynomial NARX models based on simulation error minimization},
  author={Piroddi, Luigi and Spinelli, William},
  journal={International Journal of Control},
  volume={76},
  number={17},
  pages={1767--1781},
  year={2003},
  publisher={Taylor \& Francis}
}

@article{ahmed2021stochastic,
  title={Stochastic identification of guided wave propagation under ambient temperature via non-stationary time series models},
  author={Ahmed, Shabbir and Kopsaftopoulos, Fotis},
  journal={Sensors},
  volume={21},
  number={16},
  pages={5672},
  year={2021},
  publisher={MDPI}
}

@article{proctor2016dynamic,
  title={Dynamic mode decomposition with control},
  author={Proctor, Joshua L and Brunton, Steven L and Kutz, J Nathan},
  journal={SIAM Journal on Applied Dynamical Systems},
  volume={15},
  number={1},
  pages={142--161},
  year={2016},
  publisher={SIAM}
}

@article{muld2012flow,
  title={Flow structures around a high-speed train extracted using proper orthogonal decomposition and dynamic mode decomposition},
  author={Muld, Tomas W and Efraimsson, Gunilla and Henningson, Dan S},
  journal={Computers \& Fluids},
  volume={57},
  pages={87--97},
  year={2012},
  publisher={Elsevier}
}

@article{pan2011dynamical,
  title={Dynamical mode decomposition of Gurney flap wake flow},
  author={Pan, Chong and Yu, Dongsheng and Wang, Jinjun},
  journal={Theoretical and Applied Mechanics Letters},
  volume={1},
  number={1},
  pages={012002},
  year={2011},
  publisher={Elsevier}
}

@misc{vaswani2023attentionneed,
      title={Attention Is All You Need}, 
      author={Ashish Vaswani and Noam Shazeer and Niki Parmar and Jakob Uszkoreit and Llion Jones and Aidan N. Gomez and Lukasz Kaiser and Illia Polosukhin},
      year={2023},
      eprint={1706.03762},
      archivePrefix={arXiv},
      primaryClass={cs.CL},
      url={https://arxiv.org/abs/1706.03762}, 
}

@inproceedings{zhou2021informer,
  title     = {Informer: Beyond Efficient Transformer for Long Sequence Time-Series Forecasting},
  author    = {Zhou, Haoyi and Zhang, Shanghang and Peng, Jieqi and Zhang, Shuai and Li, Jianxin and Xiong, Hui and Zhang, Wancai},
  booktitle = {Proceedings of the 35th AAAI Conference on Artificial Intelligence},
  volume    = {35},
  number    = {12},
  pages     = {11106--11115},
  year      = {2021},
  publisher = {Association for the Advancement of Artificial Intelligence},
  doi       = {10.1609/aaai.v35i12.17325}
}

@misc{wu2022autoformerdecompositiontransformersautocorrelation,
      title={Autoformer: Decomposition Transformers with Auto-Correlation for Long-Term Series Forecasting}, 
      author={Haixu Wu and Jiehui Xu and Jianmin Wang and Mingsheng Long},
      year={2022},
      eprint={2106.13008},
      archivePrefix={arXiv},
      primaryClass={cs.LG},
      url={https://arxiv.org/abs/2106.13008}, 
}

@inproceedings{
nie2023a,
title={A Time Series is Worth 64 Words:  Long-term Forecasting with Transformers},
author={Yuqi Nie and Nam H Nguyen and Phanwadee Sinthong and Jayant Kalagnanam},
booktitle={The Eleventh International Conference on Learning Representations },
year={2023},
url={https://openreview.net/forum?id=Jbdc0vTOcol}
}

@misc{loshchilov2019decoupledweightdecayregularization,
      title={Decoupled Weight Decay Regularization}, 
      author={Ilya Loshchilov and Frank Hutter},
      year={2019},
      eprint={1711.05101},
      archivePrefix={arXiv},
      primaryClass={cs.LG},
      url={https://arxiv.org/abs/1711.05101}, 
}

@misc{loshchilov2017sgdrstochasticgradientdescent,
      title={SGDR: Stochastic Gradient Descent with Warm Restarts}, 
      author={Ilya Loshchilov and Frank Hutter},
      year={2017},
      eprint={1608.03983},
      archivePrefix={arXiv},
      primaryClass={cs.LG},
      url={https://arxiv.org/abs/1608.03983}, 
}

@article{Lee2012,
  title = {Discrete-time realization of transcendental impedance models,  with application to modeling spherical solid diffusion},
  volume = {206},
  ISSN = {0378-7753},
  url = {http://dx.doi.org/10.1016/j.jpowsour.2012.01.134},
  DOI = {10.1016/j.jpowsour.2012.01.134},
  journal = {Journal of Power Sources},
  publisher = {Elsevier BV},
  author = {Lee,  James L. and Chemistruck,  Andrew and Plett,  Gregory L.},
  year = {2012},
  month = may,
  pages = {367–377}
}

@article{sancarlos2021rom,
  title={From ROM of Electrochemistry to AI-Based Battery Digital and Hybrid Twin: A. Sancarlos et al.},
  author={Sancarlos, Abel and Cameron, Morgan and Abel, Andreas and Cueto, Elias and Duval, Jean-Louis and Chinesta, Francisco},
  journal={Archives of Computational Methods in Engineering},
  volume={28},
  number={3},
  pages={979--1015},
  year={2021},
  publisher={Springer}
}

@article{li2013reduced,
  title={Reduced order model (ROM) of a pouch type lithium polymer battery based on electrochemical thermal principles for real time applications},
  author={Li, Xueyan and Xiao, Meng and Choe, Song-Yul},
  journal={Electrochimica Acta},
  volume={97},
  pages={66--78},
  year={2013},
  publisher={Elsevier}
}

\end{document}